\documentclass[prl,aps,showpacs,floatfix]{revtex4}
\usepackage{epsfig} \usepackage{graphics} \usepackage{bm}
\usepackage{amssymb}
\usepackage{graphicx}
\begin{document}

\preprint{Nata-PRB1}

\title{Quantum Theory of Lee-Naughton-Lebed's Angular Effect
in Strong Electric Fields}

\author{A.G. Lebed$^*$}

\affiliation{Department of Physics, University of Arizona, 1118 E.
4-th Street, Tucson, AZ 85721, USA}

\begin{abstract}
Some time ago, Kobayashi et al. experimentally studied the
so-called Lee-Naughton-Lebed's (LNL) angular effect in strong
electric fields [K. Kobayashi, M. Saito, E. Omichi, and T. Osada,
Phys. Rev. Lett. \textbf{96}, 126601 (2006)]. They found that
strong electric fields split the LNL conductivity maxima in
$\alpha$-(ET)$_2$-based organic conductor and hypothetically
introduced the corresponding equation for conductivity. In this
Letter, for the first time we suggest quantum mechanical theory of
the LNL angular oscillations in moderately strong electric fields.
In particular, we demonstrate that the obtained by us approximate
theoretical formula coincides with the hypothetical one and well
describes the above mentioned experiments.
\end{abstract}



\maketitle

It is well known that organic conductors, having
quasi-one-dimensional (Q1D) pieces of the Fermi surfaces (FS's),
demonstrate unique magnetic properties due to the Bragg
reflections of moving electrons from the Brillouin zones
boundaries in moderate and strong magnetic fields [1-5]. Among
them, are Field-Induced Spin(Charge)-Density-Wave (FIS(C)DW) phase
diagrams [3-15], 3D Quantum Hall Effect (3D QHE) [14-16], the
so-called Lebed's Magic Angles (LMA) [17-40], the
Lee-Naughton-Lebed's (LNL) angular oscillations [41-47], and some
others. Note the LMA phenomena [17-40] seem to be very complicated
and in most cases possess some non Fermi liquid (FL) properties
[27,29,1], whereas the FIS(C)DW, 3D QHE, and LNL phenomena were
successfully explained in the framework of the Landau FL approach
[1,2]. In particular, the LNL phenomenon was successfully
theoretically explained in Refs. [48-54]. Indeed, in Refs. [48-54]
there was considered layered Q1D conductor with the electron
spectrum,
\begin{equation}
\epsilon^{\pm}_0({\bf p})= \pm v_F (p_x \mp p_F) + 2 t_b \cos(p_y
b^*) + 2 t_c \cos(p_z c^*),
\end{equation}
where $v_F p_F \gg t_b \gg t_c$, in an inclined magnetic field,
\begin{equation}
{\bf H} = H \ (\sin \theta \cos \phi, \sin \theta \sin \phi, \cos
\theta)
\end{equation}
(see Fig.1). Note that in Eq.(1) the upper sign stands for the
right piece of the Q1D FS and the lower sign stands for the left
one. In the quasi-classical approximation, the following
expression for the LNL conductivity was derived by several
methods:
\begin{equation}
\sigma_{zz}(H, \theta, \phi) = \sigma_{zz}(0)
\sum_{N=-\infty}^{\infty}
\frac{J^2_N[\omega_c^*(\theta,\phi)/\omega_b(\theta)]}{1 + \tau^2
[\omega_c(\theta,\phi) -N \omega_b(\theta)]^2}.
\end{equation}
Note that in Eq.(3) the so-called electron cyclotron frequencies
can be expressed as [54]:
\begin{eqnarray}
\omega_b(\theta)= \frac{ev_F H b^* \cos \theta}{c}, \ \ \
\omega_c(\theta,\phi)= \frac{ev_F H c^* \sin \theta \sin \phi}{c},
\\
\omega^*_c(\theta,\phi)= \frac{ev^0_y H c^* \sin \theta \cos
\phi}{c}, \ \ \ v^0_y=2t_b b^*.
\end{eqnarray}
More recently, Kobayashi et al. [55] experimentally studied the
LNL phenomenon in rather strong electric fields and found that the
strong electric field splits the LNL maxima of conductivity (3).
What is also important they suggested a hypothetical theoretical
formula which described the above mentioned experimental
splitting.

The goal of our paper is to derive the quasi-classical expression
for conductivity in moderately strong electric and strong magnetic
fields which describes the experimentally observed splitting of
the LNL maxima of conductivity [55]. In particular, we show that
our equation has a limited area of applicability and is not
applicable in very strong electric fields.

First, let us perform the quasi-classical Peierls substitution
[56,57] for motion along the conducting chains in Eq.(1),
\begin{equation}
p_x \mp p_F = -i \frac{d}{dx},
\end{equation}
in the absence of both magnetic and electric fields:
\begin{equation}
\hat \epsilon^{\pm}_0 (x,p_y,p_z) = \mp i v_F \frac{d}{dx} + 2 t_b
\cos(p_y b^*) + 2 t_c \cos(p_z c^*).
\end{equation}
The solution of the corresponding Schr{\"o}dinger equation is
\begin{equation}
\Psi^{\pm}_0(x,p_y,p_z) = \exp \biggl(\pm i \frac{\epsilon x}{v_F}
\biggl) \exp \biggl[ \mp i \frac{2t_b x}{v_F} \cos(p_y b^*)
\biggl] \exp \biggl[ \mp i \frac{2t_c x}{v_F} \cos(p_z c^*)
\biggl],
\end{equation}
where energy $\epsilon$ is counted from the Fermi level,
$\epsilon_F = p_F v_F$.

Then we introduce the electric field applied along the least
conducting ${\bf z}$ axis as a small perturbation to the
Hamiltonian (7),
\begin{equation}
\delta \hat \epsilon (z) = eEz \ ,
\end{equation}
and perform the one more quasi-classical Peierls substitution
[56,57]:
\begin{equation}
\delta \hat \epsilon (p_z) = eEz = -ieE \frac{d}{d p_z}.
\end{equation}
In this case application of the perturbation (10) to the free
electron wave function (8) gives
\begin{equation}
\delta \hat \epsilon(p_z) \Psi^{\pm}_0(x,p_y,p_z) = \pm
\frac{eEx}{v_F} 2t_c c^* \sin(p_zc^*) \Psi^{\pm}_0 (x,p_y,p_z).
\end{equation}
It is easy to prove that for not extremely strong electric fields
the total Hamiltonian in the electric field can be written as
\begin{equation}
\hat \epsilon^{\pm} (x,p_y,p_z)= \mp i v_F \frac{d}{dx} + 2 t_b
\cos(p_y b^*) + 2 t_c \cos \biggl(p_z
c^*\mp\frac{eEc^*x}{v_F}\biggl).
\end{equation}
\begin{figure}[t]
\centering
\includegraphics[width=0.5\textwidth]{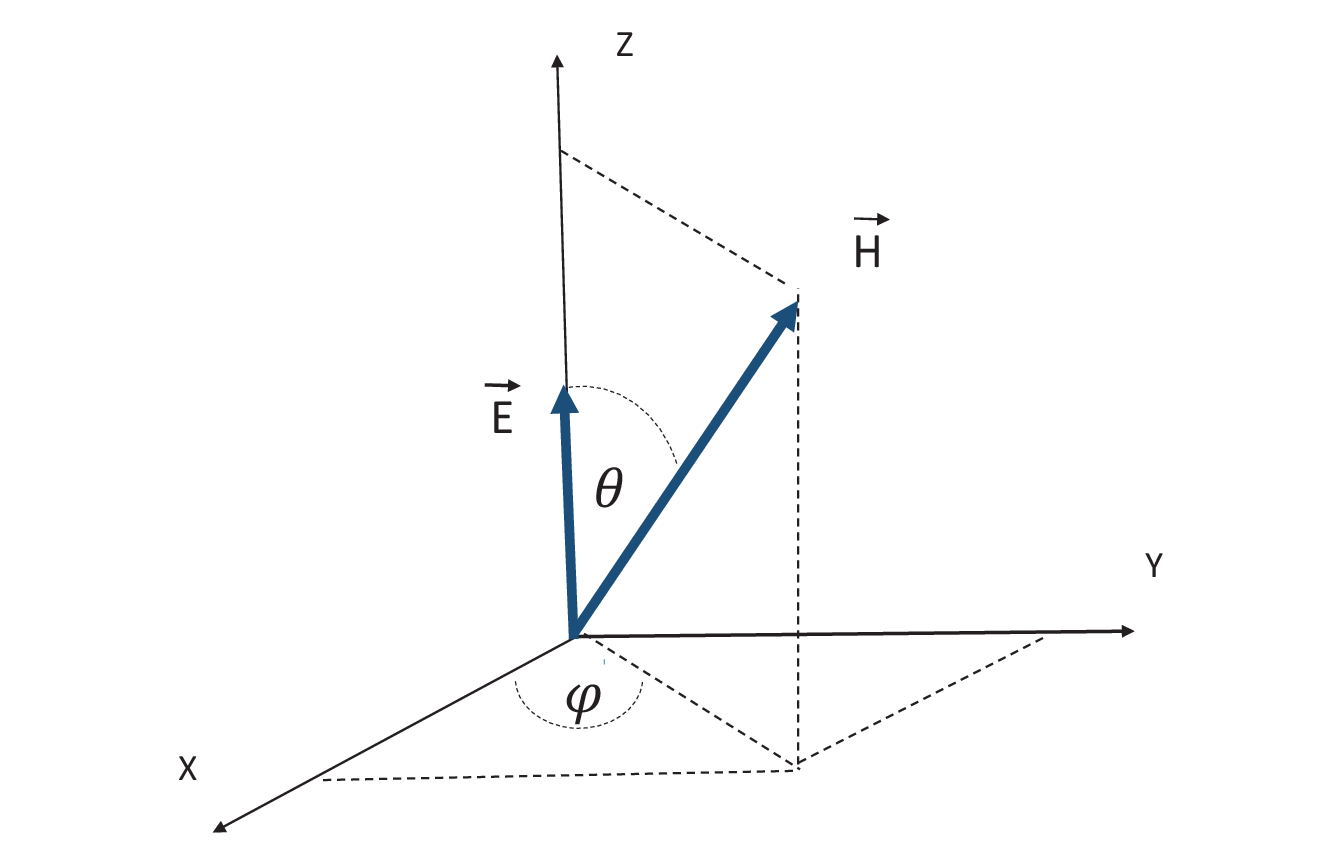}
\caption{Definition of the azimuthal angle $\theta$ and polar
angle $\phi$ for the typical Lee-Naughton-Lebed's experiment,
where $\bf z$ is the least conducting axis.}
\end{figure}

Here, we introduce the magnetic field (2) in the electron
Hamiltonian and the electron velocity operator along ${\bf z}$
axis. For the further development, it is convenient to choose
vector potential of the magnetic field in the following form:
\begin{equation}
{\bf A} = (0, x \cos \theta, -x \sin \theta \sin \phi + y \sin
\theta \cos \phi) \ H.
\end{equation}
To define the corresponding electron wave functions for the case,
where $t_b \gg t_c$, as shown in Ref.[51], it is necessary to take
into account only two first terms in Hamiltonian (12) and to
perform in the second term the following quasi-classical Peierls
substitution,
\begin{equation}
p_y \rightarrow p_y - \frac{e}{c} A_y .
\end{equation}
In this case wave function in the mixed $(x,p_y)$ representation
obeys the following Schr{\"o}dinger equation [3,51]:
\begin{equation}
\biggl(\mp i v_F \frac{d}{dx} +2t_b \cos \biggl[p_y b^*
-\frac{\omega_b(\theta)x}{v_F} \biggl]\bigg)
\Phi^{\pm}_{\epsilon}(x,p_y)=\epsilon
\Phi^{\pm}_{\epsilon}(x,p_y),
\end{equation}
where the two wave functions (8) and (15) are related by the
following equation:
\begin{equation}
\Psi^{\pm}_{\epsilon}(x,p_y)= \exp(\pm i p_F x)
\Phi^{\pm}_{\epsilon}(x,p_y).
\end{equation}
It is important that Eq.(15) can be exactly solved,
\begin{equation}
\Phi^{\pm}_{\epsilon}(x,p_y)= \exp \biggl( \pm i
\frac{\epsilon}{v_F} x \biggl) \exp \biggl\{ \pm
\frac{2it_b}{\omega_b(\theta)} \biggl( \sin
\biggl[p_yb^*-\frac{\omega_b(\theta)x}{v_F} \biggl] -\sin[p_yb^*]
\biggl) \biggl\} .
\end{equation}

Let us apply the quasi-classical Peierls substitution to energy
dependence (12) on momentum component along ${\bf z}$ axis:
\begin{equation}
\hat \epsilon_z^{\pm}(x,y,p_z) = 2 t_c \cos \biggl(p_z
c^*\mp\frac{eEc^*x}{v_F}\biggl) \rightarrow 2 t_c \cos \biggl[p_z
c^*\mp\frac{eEc^*x}{v_F} +\frac{\omega_c(\theta,\phi) x}{v_F}
-\frac{\omega^*_c(\theta,\phi) y}{v^0_y}\biggl].
\end{equation}
Taking into account that in the quasi-classical approximation
\begin{equation}
\hat v^{\pm}_z(x,y,p_z)= d[\hat \epsilon_z^{\pm}(x,y,p_z)]/dp_z, \
\ \ y=i(d/dp_y),
\end{equation}
it is possible to write the velocity component operator along
${\bf z}$ axis in the form:
\begin{equation}
\hat v^{\pm}_z(x,y,p_z) = -2 t_c c^* \sin \biggl[p_z
c^*\mp\frac{eEc^*x}{v_F} +\frac{\omega_c(\theta,\phi) x}{v_F} -i
\frac{\omega^*_c(\theta,\phi)(d/dp_y) }{v^0_y}\biggl].
\end{equation}
In Eq.(20) for the further development, we introduce
\begin{equation}
\omega^{\pm}_c(\theta,\phi)  =  \omega_c(\theta,\phi) \mp eEc^*.
\end{equation}

It is important that wave functions (17) are eigenfunctions of
velocity operator along ${\bf z}$ axis (20),(21) with the
following eigenvalues:
\begin{eqnarray}
\hat v^{\pm}_z(x,y,p_z)\Phi^{\pm}_{\epsilon}(x,p_y) = -2 t_c c^*
\sin \biggl\{p_z c^* +\frac{\omega^{\pm}_c(\theta,\phi) x}{v_F}
\pm \frac{\omega^*_c(\theta,\phi)}{\omega_b (\theta)}
\nonumber\\
\times \biggl( \cos \biggl[p_y
b^*-\frac{\omega_b(\theta)x}{v_F}\biggl] -\cos[p_yb^*] \biggl)
\biggl\} \Phi^{\pm}_{\epsilon}(x,p_y).
\end{eqnarray}
Let us apply the Kubo formula for conductivity [58,51]. We can do
this because the electron wave functions (17) and the eigenvalues
of velocity operators (22) are known. The total conductivity along
${\bf z}$ axis can be represented as a summation of the following
two contributions: one from the right sheet of the FS (1) and
another from the left sheet,
\begin{equation}
\sigma_{zz}(H,\theta,\phi)= \sigma^+_{zz}(H,\theta,\phi) +
\sigma^-_{zz}(H,\theta,\phi).
\end{equation}
By means of the Kubo formalism [58,51] we obtain
\begin{eqnarray}
\sigma^{\pm}_{zz}(H,\theta,\phi) \sim &&\int^{\pi}_{-\pi}
d(p_yb^*) \int_0^{\infty} dx \ \exp \biggl(-\frac{x}{v_F \tau}
\biggl)
\nonumber\\
&&\times \cos \biggl\{\frac{\omega^{\pm}_c(\theta,\phi) x}{v_F}
\pm \frac{\omega^*_c(\theta,\phi)}{\omega_b (\theta)} \biggl( \cos
\biggl[p_yb^*-\frac{\omega_b(\theta)x}{v_F}\biggl] -\cos[p_yb^*]
\biggl) \biggl\},
\end{eqnarray}
where $\tau$ is an electron relaxation time. Complicated double
integration in Eq.(24) can be simplified using definitions of the
Bessel functions of the N-th order, $J_N(x)$ [59,51],
\begin{equation}
\sigma^{\pm}_{zz}(H, \theta, \phi) = \frac{\sigma_{zz}(0)}{2}
\sum_{N=-\infty}^{\infty}
\frac{J^2_N[\omega_c^*(\theta,\phi)/\omega_b(\theta)]}{1 + \tau^2
[\omega^{\pm}_c(\theta,\phi) -N \omega_b(\theta)]^2},
\end{equation}
where $\sigma_{zz}(0)$ - conductivity along ${\bf z}$ axis in low
electric fields in the absence of the magnetic field. If we make
use of the Eq.(23), we finally obtain for the total conductivity
in moderately strong electric fields in the presence of the
inclined magnetic field (2):
\begin{equation}
\sigma_{zz}(H, \theta, \phi) = \frac{\sigma_{zz}(0)}{2}
\sum_{N=-\infty}^{\infty} \biggl\{
\frac{J^2_N[\omega_c^*(\theta,\phi)/\omega_b(\theta)]}{1 + \tau^2
[\omega^{+}_c(\theta,\phi) -N \omega_b(\theta)]^2} +
\frac{J^2_N[\omega_c^*(\theta,\phi)/\omega_b(\theta)]}{1 + \tau^2
[\omega^{-}_c(\theta,\phi) -N \omega_b(\theta)]^2} \biggl\}.
\end{equation}

We stress that Eq.(26) is the main result of our Letter, whereas
in Ref.[55] this equation was just guessed. Moreover, we have
shown that it is not exact and has to be used for not too high
(i.e., moderately high) electric fields. Indeed, let us discuss
its applicability. We recall that we have derived Eq.(26) using
some approximation: we have suggested that we can use Eq.(12),
instead of Eq.(11). It is easy to prove that this can be done
under the condition that
\begin{equation}
\frac{eEc^*x_0}{v_F} \ll 1 \ ,
\end{equation}
where $x_0$ is characteristic length where the integral (24)
converges. Since, as follows from (24), $x_0 \simeq v_F \tau$ the
condition (27) can be written as
\begin{equation}
eEc^* \ll 1/\tau.
\end{equation}
If we take the lowest experimentally used electric field, $V_0=Ed
=2$V, $d=0.2$mm [55], and $\hbar/\tau=$2K [1], we obtain the
inequality (28) in the form
\begin{equation}
0.25 K \ll 2 K,
\end{equation}
which shows that at lowest voltages analysis [55] is correct,
whereas at higher experimental voltages, $V_0=20$V [55], Eq.(26)
must be used with a great caution, since Eq.(28) gives quantities
of the same orders of magnitudes for the left side and for the
write one.

Let us briefly discuss one important consequence of Eq.(26) - the
splitting of the LNL maxima of conductivity in moderately strong
electric field [55]. In the limit of zero electric field at the
following typical experimental conditions, where
\begin{equation}
\omega_b(\theta) \tau \gg 1, \ \ \ \omega_c(\theta,\phi) \tau \gg
1,
\end{equation}
maxima of conductivity, as follows from Eq.(3), appear at
\begin{equation} \omega_c(\theta,\phi)= N \omega_b(\theta),
\end{equation}
where $N$ is an arbitrary integer. Under the experimental
condition (30), Eq.(26) splits each maximum into two ones which
are defined by the following equations
\begin{equation}
\omega_c(\theta,\phi)= N \omega_b(\theta) \mp  \omega_E , \
\omega_E = eEc^*.
\end{equation}
The effect of splitting was experimentally observed in Ref.[55].
Our analysis of the applicability of Eq.(26), as we discussed
above, has shown that Eqs. (32) are valid for lower experimentally
used voltages, $V_0 \simeq 2$V, and become controversial at higher
ones, $V_0 \simeq 20$V.

 The author is thankful to N.N.
Bagmet (Lebed) for useful discussions.

$^*$Also at: L.D. Landau Institute for Theoretical Physics, RAS, 2
Kosygina Street, Moscow 117334, Russia.


\begin{references}

\bibitem{Lebed-1}
A.G. Lebed ed., {\it The Physics of Organic Superconductors and
Conductors} (Springer-Verlag, Berlin, 2008).

\bibitem{Yamaji-1}
T. Ishiguro, K. Yamaji, and G. Saito, {\it Organic
Superconductors}, 2nd edn. (Springer, Berlin, 1998).

\bibitem{Lebed-2}
L.P. Gor'kov and A.G. Lebed, J. Phys. (Paris) Lett. \textbf{45},
L-433 (1984).

\bibitem{Heritier-1}
M. Heritier, G. Montambaux, and P. Lederer, J. Phys. (Paris) Lett.
\textbf{45}, L-943 (1984).

\bibitem{Chaikin-1}
P.M. Chaikin, Phys. Rev. B \textbf{31}, 4770 (1985).


\bibitem{Chaikin-1}
P.M. Chaikin, Mu-Yong Choi, J.F. Kwak, J.S. Brooks, K.P. Martin,
M.J. Naughton, E.M. Engler, and R.L. Greene, Phys. Rev. Lett.
\textbf{51}, 2333 (1983).

\bibitem{Ribault}
M. Ribault, D. Jerome, J. Tuchendler, C. Weyl, and K. Bechgaard,
J. Phys. (Paris) Lett. \textbf{44}, L-953 (1983).



\bibitem{Lebed-3}
 A.G. Lebed,
Phys. Rev. Lett. \textbf{88}, 177001 (2002).


\bibitem{Montambaux}
D. Zanchi, A. Bjelis, and G. Montambaux, Phys. Rev. B \textbf{53},
1240 (1996).

\bibitem{Brooks-1}
J.S. Qualls, L. Balicas, J.S. Brooks, N. Harrison, L.K.
Montgomery, and M. Tokumoto, Phys. Rev. B \textbf{62}, 10008
(2000).

\bibitem{Karts-1}
D. Andres, M.V. Kartsovnik, W. Biberacher, H. Weiss,
E.Balthes, H. Muller, and N. Kushch, Phys. Rev. B \textbf{64},
161104(R) (2001).

\bibitem{Karts-2}
D. Andres, M.V. Kartsovnik, P.D. Grigoriev, W. Biberacher, and H.
Muller, Phys. Rev. B \textbf{68}, 201101(R) (2003).

\bibitem{Lebed-4}
 A.G. Lebed, JETP Lett. \textbf{78}, 138 (2003).

\bibitem{Chaikin-1}
S. T. Hannahs, J. S. Brooks, W. Kang, L. Y. Chiang, and P. M.
Chaikin, Phys. Rev. Lett. \textbf{63}, 1988 (1989).

\bibitem{Cooper}
J. R. Cooper, W. Kang, P. Auban, G. Montambaux, D. Jerome, and K.
Bechgaard, Phys. Rev. Lett. \textbf{63}, 1984 (1989).

\bibitem{Yakovenko}
V. M. Yakovenko, Phys. Rev. B \textbf{43}, (1991).


\bibitem{Lebed-3} A.G. Lebed and Per Bak, Phys. Rev. Lett. \textbf{63},
1315 (1989).



\bibitem{Naughton-1}
M.J. Naughton, O.H. Chung, L.Y. Chiang, and J.S. Brooks, Material
Research Society Symposium Proceedings, \textbf{173}, 257 (1990).

\bibitem{Osada-1}
T. Osada, A. Kawasumi, S. Kagoshima, N. Miura, and G. Saito, Phys.
Rev. Lett. \textbf{66}, 1525 (1991).

\bibitem{Boebinger} G. S. Boebinger, G. Montambaux, M. L. Kaplan, R. C. Haddon, S.
V. Chichester, L. Y. Chiang, Phys. Rev. Lett. \textbf{64}, 591
(1990).

\bibitem{Naughton-2}
M. J. Naughton, O. H. Chung, M. Chaparala, X. Bu, P. Coppens,
Phys. Rev. Lett. \textbf{67}, 3712 (1991).

\bibitem{Chaikin-2}
W. Kang, S. T. Hannahs, and P. M. Chaikin, Phys. Rev. Lett.
\textbf{69}, 2827 (1992).

\bibitem{Karts-1}
M.V. Kartsovnik, A.E. Kovalev, V.N. Laukhin, and S.I. Pesotskii,
J. Phys. I (France) \textbf{2}, 223 (1992).

\bibitem{Karts-2}
M.V. Kartsovnik, A.E. Kovalev, and N.D. Kushch, J. Phys. I
(France) \textbf{3}, 1187 (1993).

\bibitem{Benhia}
K. Benhia, M. Ribault, and C. Lenior, Europhys. Lett. \textbf{25},
285 (1994).

\bibitem{Karts-2}
M.V. Kartsovnik, A.E. Kovalev, V.N. Laukhin, H. Ito, T. Ishiguro,
N.D. Kushch, H. Anzai, and G. Saito, Synth. Met. \textbf{70}, 819
(1995).

\bibitem{Chaikin-3}
E.I. Chashechkina and P.M. Chaikin, Phys. Rev. Lett. \textbf{80},
2181 (1998).

\bibitem{Osada-2}
T. Osada, H. Nose, and Kuraguchi, Physica B \textbf{294-295}, 402
(2001).

\bibitem{Chaikin-4}
E.I. Chashechkina and P.M. Chaikin, Phys. Rev. B \textbf{65},
012405 (2002).

\bibitem{Kang-1}
H. Kang, Y.J. Jo, S. Uji, and W. Kang, Phys. Rev. B \textbf{68},
132508 (2003).

\bibitem{Kang-2}
H. Kang, Y.J. Jo, and W. Kang, Phys. Rev. B \textbf{69}, 033103
(2004).

\bibitem{Ito-1}
H. Ito, D. Suzuki, Y. Yokochi, S. Kuroda, M. Umemiya, H. Miyasaka,
K-I. Sugiura, M. Yamashita, H. Tajima, Phys. Rev. B \textbf{71},
212503 (2005).

\bibitem{Karts-3}
M. V. Kartsovnik, D. Andres, S. V. Simonov, W. Biberacher, I.
Sheikin, N. D. Kushch, and H. Miller, Phys. Rev. Lett.
\textbf{96}, 166601 (2006).

\bibitem{Hill-1}
S. Takahashi, A. Betancur-Rodiguez, S. Hill, S. Takasaki, J.
Yamada, and H. Anzai, J. Low Temp. Phys. \textbf{142}, 315 (2007).

\bibitem{Kang-3}
W. Kang, T. Osada, Y.J. Jo, and Haeyong Kang, Phys. Rev. Lett.
\textbf{99}, 017002 (2007).

\bibitem{Kang}
W. Kang, Phys. Rev. B \textbf{76}, 193103 (2007).

\bibitem{Singleton-1}
A.F. Bangura, P.A. Goddard, J. Singleton, S.W. Tozer, A.I. Coldea,
A. Ardavan, R.D. McDonald, S.J. Blundell and J.A. Schlueter, Phys.
Rev. B \textbf{76}, 0525010 (2007).

\bibitem{Kang-5}
W. Kang, Ok-Hee Chung, Phys. Rev. B \textbf{79}, 045115 (2009).

\bibitem{Brooks-1}
D. Graf, J.S. Brooks, E.S. Choi, M. Almeida, R.T. Henriques, J.C.
Dias, and S. Uji, Phys. Rev. B \textbf{80}, 155104 (2009).

\bibitem{Kang-6}
W. Kang, Y.J. Jo, D.Y. Noh, K.I. Son, and Ok-Hee Chung, Phys. Rev.
B \textbf{80}, 155102 (2009).



\bibitem{Naughton-4}
M.J. Naughton, I.J. Lee, P.M. Chaikin, and G.M. Danner, Synth.
Metals \textbf{85}, 1481 (1997).

\bibitem{Yoshino}
H. Yoshino, K. Saito, H. Nishikawa, K. Kikuchi, K. Kobayashi, and
I. Ikemoto, J. Phys. Soc. Jpn. \textbf{66}, 2248 (1997).

\bibitem{Lee-1}
I.J. Lee and M.J. Naughton, Phys. Rev. B \textbf{57}, 7423 (1998).

\bibitem{Lee-2}
I.J. Lee and M.J. Naughton, Phys. Rev. B \textbf{58}, R13343
(1998).

\bibitem{Lebed-6}
A.G. Lebed, Heon-Ick Ha, and M.J. Naughton, Phys. Rev. B
\textbf{71}, 132504 (2005).

\bibitem{Lebed-7}
H.I. Ha, A.G. Lebed, and M.J. Naughton, Phys. Rev. B \textbf{73},
033107 (2006).

\bibitem{Lebed-3}
S. Wu and A.G. Lebed, Phys. Rev. B \textbf{82}, 075123 (2010).

%\bibitem{Lebed-4} A.G. Lebed, N.N. Bagmet, and M.J. Naughton, Phys.
%Rev. Lett. \underline{93}, 157006 (2004).



\bibitem{McKenzie}
R.H. McKenzie and P. Moses, Phys. Rev. B \textbf{60}, R11241
(1999).

\bibitem{Lebed-9}
A.G. Lebed and M.J. Naughton, J. Phys. IV (France) \textbf{12},
369 (2002).

\bibitem{Osada-2}
T. Osada, Physica E \textbf{12}, 272 (2002).

\bibitem{Lebed-5}
A.G. Lebed and M.J. Naughton, Phys. Rev. Lett. \textbf {91},
187003 (2003).

\bibitem{Osada-3}
T. Osada and M. Kuraguchi, Synth. Met. \textbf{133-134}, 75
(2003).

\bibitem{Yakovenko-1}
A. Banerjee and V. Yakovenko, Phys. Rev. B \textbf{78}, 125404
(2008).

\bibitem{Yakovenko-2}
B. K. Cooper and V. M. Yakovenko, Phys. Rev. Lett. \textbf{96},
037001 (2006).





\bibitem{Osada}
K. Kobayashi, M. Saito, E. Omichi, and T. Osada, Phys. Rev. Lett.
\textbf{96}, 126601 (2006).






\bibitem{Abrikosov-1}
A.A. Abrikosov, {\it Fundamentals of Theory of Metals} (Elsevier
Science, Amsterdam, 1988).

\bibitem{Lifshits}
I. M. LIfshits,  M. Ya. Azbel and M. I. Kaganov {\it Electron
Theory of Metals} (Consultants Bureau, New York, 1973).

\bibitem{Kubo}
G. Grosso and G.P. Parravichini, {\it Solid State Physics}
(Academic Press, New York, 2000).

\bibitem{GR}
L.S. Gradshteyn and I.M. Ryzhik, {\it Tables of Integrals, Series,
and Products} (5th Edn., Academic Press, Inc., London, 1994).


%\bibitem{Cooper}
%A.J. Schofield and J.R. Cooper, Phys. Rev. B \textbf{62}, 10779
%(2000).



%\bibitem{Lebed-6}
%A.G. Lebed, Phys. Rev. Lett. \textbf{95}, 247003 (2005).





\end{references}
\end{document}